\documentstyle[prl,aps,twocolumn,amstex,epsfig]{revtex}

\newcommand{\ket}[1]{\left|#1\right>} 
\newcommand{\bra}[1]{\left<#1\right|}

\newcommand{\nn}{\nonumber\\} 
 
\newcommand{\f}[1]{\mbox{\boldmath$#1$}}
\newcommand{\fk}[1]{\mbox{\boldmath$\scriptstyle#1$}}

\newcommand{\na}{\mbox{\boldmath$\nabla$}}
\newcommand{\bea}{\begin{eqnarray}}
\newcommand{\ea}{\end{eqnarray}}

\begin{document} 
 
\wideabs{
\title{Resonant cavity photon creation via the dynamical Casimir effect}
\author{
Michael Uhlmann, G\"unter Plunien, Ralf Sch\"utzhold, and Gerhard Soff}
\address{Institut f\"ur Theoretische Physik,
Technische Universit\"at Dresden,
01062 Dresden, Germany}
\date{\today}
\maketitle
\begin{abstract} 
Motivated by a recent proposal for an experimental verification of
the dynamical Casimir effect, the macroscopic electromagnetic field
within a perfect cavity containing a thin slab with a time-dependent
dielectric permittivity is quantized in terms of the dual potentials. 
For the resonance case, the number of photons created out of the
vacuum due to the dynamical Casimir effect is calculated for both
polarizations (TE and TM).  
\\
PACS:
42.50.Lc, 
03.70.+k, 
42.50.Dv, 
42.60.Da. 
\end{abstract} 
}

\begin{figure}[ht]
\centerline{\mbox{\epsfxsize=6cm\epsffile{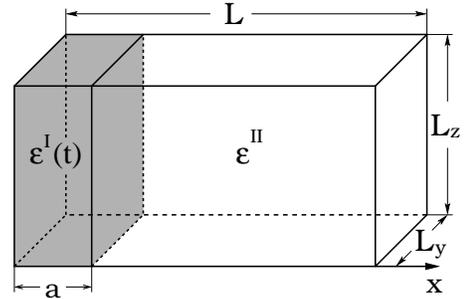}}}
\caption{Sketch of the (lossless) cavity containing a thin slab with a 
  time-dependent dielectric permittivity $\varepsilon^I(t)$.}
\label{cavity}
\end{figure}

One of the most impressive manifestations of the non-trivial structure
of the vacuum is the static Casimir effect, i.e., the attraction of
two perfectly conducting plates, for example, generated by the
corresponding distortion of the electromagnetic vacuum state
\cite{casimir}. 
The non-inertial motion of a mirror can even create particles 
(i.e., photons) out of the vacuum \cite{moore}
due to the time-dependent disturbance -- which is called (in analogy)
dynamical Casimir effect (see, e.g., \cite{dodonov} for review).   
Unfortunately, in contrast to the former (static) effect, the latter
(dynamical) prediction has not been experimentally verified yet.
To this end, it is probably advantageous to exploit the drastic
enhancement of the number of created photons within a cavity
occurring if the frequency of the wall vibration is in resonance
with one of the (discrete) cavity modes. 
The difficulty of accomplishing mechanical vibrations of the wall
with high frequencies (and appropriate amplitudes) experimentally
has lead to the idea of simulating the wall motion by manipulating
the dielectric permittivity (or magnetic permeability) of some medium 
within the cavity (which can be done much faster).
E.g., filling the whole cavity with a homogeneous medium described by
a time-dependent permittivity $\varepsilon(t)$ is analogous to
introducing an effective length of the cavity via 
$L_{\rm eff}(t)=\sqrt{\varepsilon(t)}\,L$.
However, since it is rather difficult to influence a medium filling
the complete cavity, a new proposal \cite{ruoso} (see also \cite{lozovik}) 
for an experimental verification of the dynamical Casimir effect
envisions a small slab with a fixed thickness $a$ and a time-dependent
permittivity $\varepsilon(t)$ located at one of the walls of the
cavity, cf.~Fig.~\ref{cavity}.   
The question of whether and how the motion of the cavity wall can be
simulated by such a small dielectric slab -- especially in view of the
number of created photons -- will be the subject of the subsequent
considerations (see also \cite{johnston} for a 1+1 dimensional
scalar field model). 

In this Letter, we present an {\em ab initio} derivation of the
dynamical Casimir effect based on the quantization of the full
macroscopic electromagnetic field within a (perfect) cavity with
space-time dependent dielectric properties -- superseding previous
effectively 1+1 dimensional calculations  
(scalar field model, see, e.g., \cite{johnston,eff-1+1}) 
and approaches based on special factorization assumptions 
(see, e.g., \cite{factor}).
For 3+1 dimensional cavities with moving walls, there exist various
calculations for scalar fields, but very few taking into account the
full electromagnetic field. 
E.g., in \cite{maia-neto}, the electromagnetic field is
effectively split up into two independent scalar fields obeying
different  boundary conditions via introducing different potentials
for the TE and the TM modes -- which leads to a decoupling of the 
polarizations (TE and TM) per construction. 
However, in the most general situation, TE and TM modes can mix -- 
hence their coupling should not be excluded {\em a priori} but
investigated for each special case.

Since we are considering low-frequency (e.g., microwave) photons only, 
we start from the macroscopic source-free Maxwell equations
($\varepsilon_0=\mu_0=\hbar=1$)
\bea
\label{maxwell}
\na\cdot\f{B}=\na\cdot\f{D}=0
\,,\;
\f{\dot D}=\na\times\f{H}
\,,\;
\f{\dot B}=-\na\times\f{E}
\,,
\ea
with $\f{H}(t,\f{r})=\f{B}(t,\f{r})$ and
$\f{D}(t,\f{r})=\varepsilon(t,\f{r})\f{E}(t,\f{r})$.
If we were to use the usual vector potential $\f{A}$ in temporal gauge 
($\Phi=0$) 
\bea
\label{normal}
\f{E}=\f{\dot A}
\,,\;
\f{B}=-\na\times\f{A}
\,,
\ea
the constraint $\na\cdot[\varepsilon(t,\f{r})\f{\dot A}(t,\f{r})]=0$
would render the usual canonical quantization 
\bea
\label{commutator}
\left[\hat A_i(t,\f{r}),\hat D_j(t,\f{r'})\right]=
C_{ij}(\f{r},\f{r'})
\,,
\ea
in connection with eliminating the longitudinal degree of freedom
rather tedious (cf.~also \cite{quant-rad}) because, in this case, 
$\na\cdot\f{D}=0$ implies $\partial_j'C_{ij}=0$ but
$\partial_iC_{ij}\neq0$ in general.   

Therefore, we avoid these difficulty with the well-known trick of
introducing the dual vector potential 
(see, e.g., \cite{maia-neto}) 
\bea
\label{dual}
\f{H}=\f{\dot\Lambda}
\,,\;
\f{D}=\na\times\f{\Lambda}
\,,
\ea
which applies in this form to the source-free Maxwell equations 
(\ref{maxwell}) only. 
In terms of the dual vector potential, the constraint simply reads 
$\na\cdot\f{\Lambda}=0$. 
After the duality transformation \cite{duality}, the Lagrangian is
still the usual Larmor invariant -- but with the opposite sign 
\bea
\label{lagrangian}
{\mathfrak L}
=
\frac12\int d^3r\left[\f{B}\cdot\f{H}-\f{E}\cdot\f{D}\right]
\,,
\ea
and the Hamiltonian is again the total energy
\bea
\label{hamiltonian}
{\mathfrak H}
&=&
\frac12\int d^3r\left[\f{B}\cdot\f{H}+\f{E}\cdot\f{D}\right]
\nn
&=&
\frac12\int d^3r
\left[\f{\dot\Lambda}^2+\frac1\varepsilon(\na\times\f{\Lambda})^2\right]
\,.
\ea
The continuity conditions 
($\Delta\f{\Lambda}=\f{\Lambda}^I-\f{\Lambda}^{II}$)
for the dual vector potential at the
interface between the regions $I$ and $II$ of the cavity can be
derived from the Maxwell equations (\ref{maxwell})
\bea
\label{continuity}
\Delta\f{\Lambda}
=
\Delta\left(\na\times\f{\Lambda}\right)_\perp
=
\Delta\left(\frac1\varepsilon\,\na\times\f{\Lambda}\right)_\|
=0
\,.
\ea
Assuming that the walls of the cavity are perfectly conducting, 
for example, the boundary conditions read
\bea
\label{boundary}
\f{E}_\|=0
\,\leadsto\,
(\na\times\f{\Lambda})_\|=0
\,\leadsto\,
\left(\f{\Lambda}\times[\na\times\f{\Lambda}]\right)_\perp=0
\,.
\ea
Consequently, the boundary term arising from the integration by parts 
(as in the Poynting theorem) of the term $(\na\times\f{\Lambda})^2$ in
Eq.~(\ref{hamiltonian}) vanishes.
Hence we can introduce a non-negative and self-adjoint operator 
$\cal K$ via
\bea
\label{operator}
{\cal K}\f{f}_\alpha
=
\na\times\left(\,\frac1\varepsilon\,\na\times\f{f}_\alpha\right)
=
\Omega^2_\alpha\f{f}_\alpha
\,,
\ea
with eigenfunctions $\f{f}_\alpha$ and eigenvalues $\Omega^2_\alpha$.
Note that we consider a lossless (ideal) dielectric medium resulting
in a real permittivity $\varepsilon\in\mathbb R$ 
(and hence a self-adjoint operator $\cal K$). 
Owing to the time-dependence of the dielectric permittivity
$\varepsilon(t,\f{r})$, the operator ${\cal K}(t)$ and consequently
its eigenfunctions $\f{f}_\alpha(t,\f{r})$ as well as eigenvalues 
$\Omega^2_\alpha(t)$ are also explicitely time-dependent in general.

The longitudinal modes $\f{f}_\alpha^\|$ form the (orthogonal)
eigenspace with zero eigenvalue $\na\times\f{f}_\alpha^\|=0$ and hence
we can restrict the operator $\cal K$ to the constraint sub-space
$\na\cdot\f{f}_\alpha=0$. 
Since $\cal K$ is a real operator, we can choose its eigenfunctions to
be real as well $\f{f}_\alpha=\f{f}_\alpha^*$; and because $\cal K$ is 
self-adjoint, its eigenfunctions are orthonormal (for equal times) 
\bea
\label{orthonormal}
\int d^3r\,\f{f}_\alpha(t)\cdot\f{f}_\beta(t)=\delta_{\alpha\beta}
\,,
\ea
and complete
\bea
\label{complete}
\sum\limits_{\alpha}f^i_\alpha(t,\f{r})f^j_\alpha(t,\f{r'})
=
\delta_\perp^{ij}(\f{r}-\f{r'})
\,,
\ea
with $\delta_\perp^{ij}(\f{r}-\f{r'})$ denoting the transversal Dirac 
$\delta$-distribution $\partial_i\delta_\perp^{ij}(\f{r}-\f{r'})=0$.
Hence a corresponding normal mode expansion of the Lagrangian and the
Hamiltonian in terms of the dual potentials into the instantaneous
basis
\bea
\label{instantaneous}
\f{\Lambda}(t,\f{r})=\sum\limits_{\alpha}Q_\alpha(t)\f{f}_\alpha(t,\f{r})
\,,
\ea
leads to (see also \cite{canon})
\bea
\label{canonical}
{\mathfrak H}(t)
=
\frac12\sum\limits_{\alpha}(P_{\alpha}^2+\Omega^2_\alpha(t)Q_{\alpha}^2)
+
\sum\limits_{\alpha\beta}P_{\alpha}Q_{\beta}{\cal M}_{\alpha\beta}(t)
\,.
\ea
From now on, we shall drop the summation signs for convenience by
declaring a corresponding (Einstein-like) sum convention.
The canonical conjugated momenta are given by 
$P_{\alpha}=\dot Q_{\alpha}+{\cal M}_{\alpha\beta}(t)Q_{\beta}$
and the anti-symmetric inter-mode coupling matrix reads
\bea
\label{matrix}
{\cal M}_{\alpha\beta}(t)
=
\int d^3r\,\f{f}_\alpha(t)\cdot\f{\dot f}_\beta(t)
\,.
\ea
The usual equal-time canonical commutation relations, e.g.,  
$[\hat Q_\alpha,\hat P_\beta]=i\delta_{\alpha\beta}$,
are equivalent to the commutators for the fields, such as  
$[\hat\Lambda^j(t,\f{r}),\hat\Pi^k(t,\f{r'})]
=i\delta_\perp^{jk}(\f{r}-\f{r'})$.

It will be convenient to classify the eigenmodes with respect to their 
polarization at the interface into TE (transversal electric) and TM
(transversal magnetic) modes  
\bea
\label{te-tm}
E^{TE}_\perp=0
\,,\;
B^{TM}_\perp=0
\,.
\ea
In terms of the dual potentials, these conditions read 
$(\na\times\f{\Lambda})^{TE}_\perp=0$ and $\Lambda^{TM}_\perp=0$. 
Assuming the absence of any static fields and fields outside the
cavity (the walls are supposed to be perfectly reflecting), the
boundary condition $B_\perp=0$ implies $\Lambda_\perp=0$ and hence 
Eq.~(\ref{boundary}) imposes the condition
$\nabla_\perp\f{\Lambda}_\|=0$.
As a result, we can make the following separation {\em ansatz} in the
homogeneous region $I$ 
\bea
\label{ansatz}
\f{f}_{\fk{k}\sigma}^I=
\left( 
\begin{array}{c}
\sin(k_x^I x) \cos(k_y y) \cos(k_z z) \epsilon_{\fk{k} \sigma x}^I \\
\cos(k_x^I x) \sin(k_y y) \cos(k_z z) \epsilon_{\fk{k} \sigma y}^I \\
\cos(k_x^I x) \cos(k_y y) \sin(k_z z) \epsilon_{\fk{k} \sigma z}^I \\
\end{array}
\right)
\,,
\ea
and analogously for region $II$ with $x$ being replaced by $x-L$.

The wave-numbers $k_y$ and $k_z$ are simply determined by the
perpendicular cavity dimensions $L_y$, $L_z$ via $k_y=n_y\pi/L_y$ 
and $k_z=n_z\pi/L_z$, respectively, with integers $n_y$ and $n_z$.
The remaining polarization factors $\f{\epsilon}_{\fk{k}\sigma}^{I/II}$ 
as well as the $k_x^{I/II}$-values have to be determined according to 
the continuity conditions in Eq.~(\ref{continuity}),
the polarization condition (TE or TM) in Eq.~(\ref{te-tm}),
the transversality condition $\na\cdot\f{\Lambda}=0$,
the overall normalization in Eq.~(\ref{orthonormal}), and, finally, 
the eigenvalue equation (for a fixed time)
\bea
\label{eigenvalue}
\Omega^2
=
\frac{(k_x^{I})^2+k_y^2+k_z^2}{\varepsilon^{I}}
=
\frac{(k_x^{II})^2+k_y^2+k_z^2}{\varepsilon^{II}}
\,,
\ea
which provides a relation between $k_x^I$ and $k_x^{II}$.
Using the conditions mentioned above, we arrive at the transcendental 
equations  
\bea
\label{matching}
{\rm TE} 
&:& 
\frac{\tan(ak_x^I)}{k_x^I}
=
\frac{\tan(k_x^{II}[a-L])}{k_x^{II}}
\,,
\nn
{\rm TM} 
&:& 
\frac{k_x^I\tan(ak_x^I) }{\varepsilon^I}
=
\frac{k_x^{II}\tan(k_x^{II}[a-L])}{\varepsilon^{II}}
\,,
\ea
which have to be satisfied simultaneously to the eigenvalue equation
(\ref{eigenvalue}).

Assuming the slab to be sufficiently small $a \ll L$, we can find
approximate solutions for the TE modes
\bea
\label{k2-te}
k_x^{II} 
=
\frac{n_x\pi}{L}
+{\cal O}\left(\frac{a^3}{L^3}\right)
\,,
\ea
and for the TM modes (for $n_x>0$)
\bea
\label{k2-tm}
k_x^{II} 
=
\frac{n_x\pi}{L}
\left(1+\frac{a}{L}
\left[\frac{\varepsilon^{II}}{\varepsilon^{I}}-1\right]
\frac{k_\|^2}{k_\perp^2}
\right)
+{\cal O}\left(\frac{a^2}{L^2}\right)
\,,
\ea
with $k_\|^2=k_y^2+k_z^2$ and $k_\perp=n_x\pi/L$.

We observe that the first-order (in $a/L\ll1$) contributions to the
eigenvalues $\Omega^2_\alpha$ of the TE modes are independent of 
$\varepsilon^{I/II}(t)$. 
Only for the TM modes, a variation of the permittivities
$\varepsilon^{I/II}(t)$ induces a change of the eigenvalues 
(with the label $\alpha=\{\f{n},{\rm TM}\}$) 
\bea
\label{delta-omega}
\Delta\Omega^2_{\fk{n},{\rm TM}}(t)
&=&
\frac{2}{\varepsilon^{II}}\,
k_\|^2\,
\frac{a}{L}\,
\left[\frac{\varepsilon^{II}}{\varepsilon^{I}(t)}-1\right]
+{\cal O}\left(\frac{a^2}{L^2}\right)
\,.
\ea
The first-order term of the coupling matrix can be derived in complete
analogy: ${\cal M}_{\alpha\beta}(t)\propto(a/L)\partial_t
(\varepsilon^{II}/\varepsilon^{I})$.

In order to simulate an oscillation of the wall, we assume a harmonic
time-dependence of the ratio
\bea
\label{harmonic}
\frac{\varepsilon^{II}}{\varepsilon^{I}(t)}=\xi+\chi\sin(\omega t)
\,,
\ea
with the amplitude $\chi$ and an irrelevant additive constant $\xi$
(which just induces a constant shift of the eigenfrequencies).
A small harmonic perturbation over a relatively long time duration 
(i.e., many oscillations) enables us to employ the rotating wave
approximation, which neglects all non-resonant terms.
According to Eq.~(\ref{canonical}) with 
$\Omega^2_\alpha(t)=(\Omega^0_\alpha)^2+\Delta\Omega^2_\alpha(t)$, 
the perturbation Hamiltonian can
be split up into two parts, the diagonal (so-called squeezing) term 
$\Delta\Omega^2_\alpha(t)\hat Q_{\alpha}^2/2$ and the off-diagonal 
(so-called velocity) contribution 
$\hat P_{\alpha}\hat Q_{\beta}{\cal M}_{\alpha\beta}(t)$,
cf.~\cite{canon}.
The resonance condition for the former (squeezing) term reads 
$\omega=2\Omega^0_\alpha$ and for the latter inter-mode coupling
(velocity) contribution $\omega=|\Omega^0_\alpha\pm\Omega^0_\beta|$. 
In the following, we shall assume a cavity with well-separated
eigenfrequencies where the external oscillation frequency $\omega$
matches the diagonal resonance condition $\omega=2\Omega^0_\alpha$ for
a certain TM mode only (no resonant inter-mode coupling). 
In this case, the effective Hamiltonian reads
\bea
\label{effective}
\hat{\mathfrak H}_{\rm eff}^{\rm TM}
=
\frac{i}{2}\,
\frac{k_\|^2}{\omega}\,
\frac{\chi}{\varepsilon^{II}}\,
\left([\hat a^\dagger_\alpha]^2-\hat a_\alpha^2\right)
\frac{a}{L}
+
{\cal O}\left(\frac{a^2}{L^2}\right)
\,,
\ea
for the resonant TM mode $\alpha$.
Accordingly, the time-dependence of the dielectric permittivity of the
thin slab induces the creation of an exponentially increasing number
of particles (photons) out of the vacuum (dynamical Casimir effect)
\bea
\label{particles}
\langle\hat N\rangle(t)
&=&
\bra{0}
\exp\{+i\hat{\mathfrak H}_{\rm eff}^{\rm TM}t\}\,
\hat a^\dagger\hat a\,
\exp\{-i\hat{\mathfrak H}_{\rm eff}^{\rm TM}t\}
\ket{0}
\nn
&=&
\sinh^2\left(
\frac{k_\|^2}{\omega}\,
\frac{\chi}{\varepsilon^{II}}\,
\frac{a}{L}\,t\right)
\,.
\ea
Let us state the physical assumptions entering the above derivation.
Firstly, by starting from the source-free macroscopic Maxwell
equations with perfectly conducting boundary conditions we assumed
an ideal cavity and omitted losses and decoherence etc.
Of course, the applicability of this assumption has to be checked
(e.g., whether the Q-factor of the cavity is large enough)
before conducting a corresponding experiment.
Secondly, the external oscillation was assumed to be harmonic with the
frequency matching the resonance condition exactly.
Other periodic time-dependences could lead to the contribution of
higher harmonics ($2\omega$, etc.) and one has to make sure that the
possibly resulting inter-mode coupling does not spoil the main
contribution in Eq.~(\ref{particles}).
A deviation (detuning) from the exact resonance 
$\omega=2\Omega^0_\alpha(1+\delta)$ is also not critical as long as the
relative detuning $\delta$ is smaller than the relative perturbation
amplitude, cf.~\cite{detuning}. 
Thirdly, the neglect of the higher-order terms in the Taylor expansion
of the transcendental matching equations  (\ref{matching}) leading to
Eqs.~(\ref{k2-te}) and (\ref{k2-tm}) assuming a small slab $a \ll L$ 
is only justified if all other involved quantities are not too large.
If the ratio $\varepsilon^{II}/\varepsilon^{I}(t)$ changes
drastically, this approximation breaks down as soon as the smallness
of the expansion parameter $a/L$ is compensated by a huge variation in  
$\varepsilon^{II}/\varepsilon^{I}(t)$. 

Let us study the two limiting cases:
For $\varepsilon^{I}\gg\varepsilon^{II}$, the wave-numbers
behave as $k_x^{I} \gg k_x^{II}$ according to Eq.~(\ref{eigenvalue}). 
Hence the poles of the functions $\tan(ak_x^I)$ in
Eq.~(\ref{matching}) induce drastic changes of $k_x^{II}$ and 
thus $\Omega$ -- for both, TE and TM modes.
In this case, the eigenmodes are 'pulled' into the small slab
(which is not desirable since the time-varying material properties will 
entail the danger of dissipation and decoherence).
In the opposite case $\varepsilon^{I}\ll\varepsilon^{II}$, the
wave-number in the small slab becomes imaginary,
cf.~Eq.~(\ref{eigenvalue}).  
In some sense, the modes are 'pushed' out of the small slab
similar to the phenomenon of total reflection 
(as in an optical fiber, for example) in this situation.
With exactly the same argument as before, there is no effect to lowest
order in $a/L$ for the TE modes. 
However, the values $k_x^{II}$ and hence $\Omega$ of the TM modes
change strongly owing to the occurrence of the term
$\varepsilon^{II}/\varepsilon^{I}$, cf.~Eqs.~(\ref{k2-tm}) and  
(\ref{delta-omega}).

In summary, a (small and smooth) motion of the cavity wall 
(more precisely, its resonant features) can only be simulated for TM 
modes with  
$L_{\rm eff}^{\rm TM}=L[1-(a/L)\{\varepsilon^{II}/\varepsilon^{I}-1\}
(k_\|^2/k_\perp^2)]+{\cal O}(a^2/L^2)$
via moderate changes of $\varepsilon^{II}/\varepsilon^{I}(t)$ 
-- huge variations are inappropriate.
At a first glance, this result might be a bit surprising since the
initial (e.g., $\varepsilon^{II}=\varepsilon^{I}$) and final situations
($\varepsilon^{II}$ and $\varepsilon^{I}$ vastly different) are
similar to two cavities with slightly different dimensions.
However, one has to bear in mind that the whole evolution -- and not
just the initial and final state -- is important for the dynamical
Casimir effect.  
E.g., instead of moving one wall of the cavity, one could mediate
between the same initial and final states by inserting an additional 
wall into the cavity -- which would imply a completely different
dynamics 
(e.g., depending on what happens in the cut-off part of the cavity).

With the aid of the dual vector potential $\f{\Lambda}$, we were able 
to quantize the (macroscopic) electromagnetic field within a cavity
with space-time dependent dielectric properties $\varepsilon(t,\f{r})$ 
and $\mu=\rm const$ facilitating the investigation of the influence
of the polarization (TE and TM modes) etc.   
In the opposite case $\mu(t,\f{r})$ and $\varepsilon=\rm const$, 
an analogous derivation can be accomplished using the ordinary vector
potential $\f{A}$.
In contrast to the former case, where only the TM modes feel 
(to first order) the presence of the dielectric slab, both
polarizations (TE and TM) are effected by the changing magnetic
properties $\mu(t,\f{r})$ in the latter situation \cite{prepare}.
The physical difference between the two cases can be explained by
the distinct boundary conditions which involve real macroscopic
charges and currents and are therefore not invariant under the duality 
transformation \cite{duality}. 

In conclusion, there are three major differences between the scenario
under investigation and a cavity with a moving wall: firstly, the
dependence on the polarization (only TM modes), secondly, the
dependence on the perpendicular wave-number 
$\propto k_\|^2$, and thirdly, that fact that the
effect does not vanish for $n_x=0$ -- in this case, one obtains just
half the value given in Eq.~(\ref{delta-omega}). 

Let us insert some explicite numbers: 
If we assume a switching time of about 100 ps \cite{ruoso}, the term 
$k^2_\|/\omega$ in Eq.~(\ref{particles}) is of order GHz.
According to the resonance condition, the dimensions of the cavity
and hence the wave-length of the created photons should be several
centimeters (i.e., microwaves). 
The second term $\chi/\varepsilon^{II}$ cannot exceed 1/2 and is
assumed to be of order one.
The remaining parameter $a/L$ should be small, say 1/100 -- 
where the explicite value of $a$ depends on the way of changing the
dielectric properties, e.g., laser illumination depth \cite{ruoso}.
Note that $a/L=1/100$ is still by far larger than the relative
amplitudes that can be achieved by mechanical vibrations 
(at these frequencies).
Provided that the assumption of a perfect cavity (e.g., Q-factor) is
appropriate during such time-scales, one would create a significant
amount of photons after a few microseconds. 

{\em Acknowledgments}\;
M.~U.~and R.~S.~gratefully acknowledge financial support by the
Emmy-Noether Programme of the German Research Foundation (DFG) under
grant No.~SCHU 1557/1-1.  
G.~P.~and G.~S.~acknowledge support by BMBF, DFG and GSI (Darmstadt).

\addcontentsline{toc}{section}{References}


\begin{thebibliography}{99}
\bibitem{duality}
In vacuum 
($\partial_\mu F^{\mu\nu}=\partial_\mu\widetilde F^{\mu\nu}=0$), 
the duality transformation is just $F_{\mu\nu}\to\widetilde F_{\mu\nu}=
\epsilon_{\mu\nu\rho\sigma}F^{\rho\sigma}/2$
with $F_{\mu\nu}(\f{E},\f{B})$ being the field strength tensor and  
$\epsilon_{\mu\nu\rho\sigma}$ the Levi-Civita symbol.
%
In a medium, one has to distinguish $F_{\mu\nu}(\f{E},\f{B})$ and 
$G_{\mu\nu}(\f{D},\f{H})$ with $\partial_\mu\widetilde F^{\mu\nu}=0$
and $\partial_\mu G^{\mu\nu}=0$, and hence the duality transformation
is $F_{\mu\nu}\to\widetilde G_{\mu\nu}$ as well as 
$G_{\mu\nu}\to\widetilde F_{\mu\nu}$, i.e., 
$\f{E}\to\f{H}$, 
$\f{B}\to-\f{D}$, 
$\f{D}\to\f{B}$, and
$\f{H}\to-\f{E}$.
\end{thebibliography}
\end{document}